# Atomic-scale simulations on the sliding of incommensurate surfaces: The breakdown of superlubricity

Woo Kyun Kim

*Department of Materials Science and Engineering, Johns Hopkins University, Baltimore, Maryland, 21218, USA*

Michael L. Falk

*Department of Materials Science and Engineering, Johns Hopkins University, Baltimore, Maryland, 21218, USA*

*Department of Mechanical Engineering, Johns Hopkins University, Baltimore, Maryland, 21218, USA*

*Department of Physics and Astronomy, Johns Hopkins University, Baltimore, Maryland, 21218, USA*

Molecular dynamics simulations of frictional sliding in an Atomic Force Microscope (AFM) show a clear dependence of superlubricity between incommensurate surfaces on tip compliance and applied normal force. While the kinetic friction vanishes for rigid tips and low normal force, superlubric behavior breaks down for softer tips and high normal force. The simulations provide evidence that the Frenkel-Kontorova-Tomlinson (FKT) scaling applies equally to a more realistic 3-D incommensurate AFM model except in the limit of very low stiffness and high normal load limit. Unlike the FKT model in which the breakdown of superlubricity coincides to the emergence of the meta-stable states, in the 3-D model some meta-stable states appear to reduce frictional force leading to non-monotonic dependence of force on normal load and tip compliance. Meta-stable states vary with the slider positions, and the relative stabilities of these meta-stable states result in varying transition mechanisms depending on sliding velocity.

## I. INTRODUCTION

The triobological properties of sliding surfaces result from forces acting on a large number of asperities that comprise the interface. While some phenomenological observations such as Amontons' laws are well established,[1] the fundamental understanding of friction as it arises from single-asperity contacts remains a significant challenge in part due to a lack of knowledge about the physical processes that dominate at the smallest length scales.[2] In recent years experimental advances including the invention of Atomic Force Microscope (AFM)[3] have made it possible to measure the friction force acting on a single-asperity contact at the nano-meter scale and to investigate friction on a fundamental level.[2-6] Moreover, developments in both hardware and methodologies have allowed computer simulations to play an important role in interpreting experiments and understanding the origins of energy dissipation during sliding at the nano-meter scale. [7-12]

The friction force measured in an AFM experiment arises from the interactions of atoms at the interface of the AFM tip and the scanned surface. Since most conventional AFM tips produce multiple-atom rather than single-atom contacts, the relative arrangement of these atoms at the interface significantly affects the friction force. One particularly interesting phenomenon closely related to this effect is superlubricity, dry sliding with very low kinetic friction that has been observed in experiments of incommensurate surfaces in the absence of adsorbates.[4-6] Dienwiebel *et al*.[5] measured the friction force of a crystalline graphite surface using a tungsten tip and rotated the sample surface over a range of 100 degrees. They observed nearly zero friction force over a wide range of angles (~ 60 degrees) separated by two narrow angular regions with high friction. This result implies that they actually measured the force between the graphite substrate and a graphite flake attached to the tip, and that the friction force vanishes when the crystalline surfaces are out of registry, i.e., incommensurate. Another experimental study has shown that superlubric behavior also depends on the applied normal force. In an experiment with NaCl crystals Socoliuc *et al*.[6] observed that the surfaces, which exhibited vanishing friction force at low normal loads, show high friction force at high normal loads. They conjecture that the system has only one global minimum at low normal loads so that the system undergoes continuous sliding, but the increase in the normal load creates multiple meta-stable states causing energy dissipation during the transitions between these states.

The phenomenon of vanishing friction also arises in theoretical models consisting of one-dimensional infinite atomic chains such as the Frenkel-Kontorova-Tomlinson (FKT) model.[13] In the FKT model rigid incommensurate surfaces can exhibit zero static friction with non-vanishing surface energy corrugation and zero kinetic friction in the zero velocity limit. Note that zero static friction always implies zero kinetic friction although zero kinetic friction does not necessarily imply zero static friction, and zero kinetic friction is possible only when there is one and only one stable state. Moreover, when the rigidity of surfaces is low compared to the interfacial stiffness between the surfaces, the incommensurate FKT model does not exhibit superlubric behavior anymore. A simulation study with flat crystalline walls [9] has also been shown to exhibit a similar dependence of vanishing friction on the surface compliance. In the FKT model this is accompanied by the emergence of a set of meta-stable states, but this phenomenon of emerging meta-stable states has not been directly investigated in more realistic three-dimensional systems.

While increasing attention has been devoted to superlubricity, simulation studies using realistic AFM models have not reported on this phenomenon, particularly in connection with variations in the tip compliance and the normal force. Moreover, we do not know whether or not Socoliuc's experimental observation [6], the breakdown of superlubricity with increasing normal force, is true in every superlubric system. Furthermore, as mentioned above, the emergence of meta-stable states and their effects on friction has not been directly investigated in either experiments or simulations.

In this paper we model an AFM experiment of incommensurate surfaces using molecular dynamics (MD) simulations to systematically study the effects of the critical parameters on friction: the tip compliance and the normal force. Instead of using flat surfaces [9] or a one-dimensional chain of atoms as in the FKT model,[13] a simple model of an AFM tip with a realistic 3-dimensional shape has been studied to capture sliding behaviors close to those observed in experiments.

We present the simulation results obtained by varying both tip stiffnesses and normal forces. The transition from the superlubric regime to a finite large friction force regime has been observed depending on the relative magnitudes of tip stiffnesses and normal forces. In addition to measuring the friction force, we also analyzed the system behaviors in terms of the transitions between multiple meta-stable states identified during simulations. This analysis has been designed to see whether meta-stable states emerge in the more realistic model as they do in the FKT model. In addition to the existence of meta-stable states causing the stick-slip motion in the sliding direction, the existence of meta-stable states in the stick-phase and their effect on the transition mechanisms have also been studied.

## II. MODEL

### A. A review of the FKT model

In this section, we briefly review the FKT model and discuss the limitations of the model describing the sliding behavior of incommensurate surfaces, especially in connection with AFM experiments. The MD simulation model and methods used in this study are described in Sec. II. B.

Figure 1 illustrates the Frenkel-Kontorova (FK) model,[14] the precursor of the FKT model, and the FKT model.[13] Both models have $N$ atoms interacting with a one-dimensional sinusoidal potential with an energy corrugation of $b$ and a periodicity of $c_1$. The atoms can also interact with each other by a harmonic potential with a spring constant $k_1$ and an equilibrium length of $c_2$. Unlike the Tomlinson model [15, 16] driven by an external spring connected to a sliding body, the FK model cannot describe driven systems. To overcome this disadvantage, in the FKT model the atoms are coupled to a rigid sliding body as shown in Fig. 1 (b) so that the system can be driven by applying a constant force or a constant velocity to the rigid body. The atoms are coupled to fixed points in the rigid body, separated by $c_2$, through a harmonic potential with a spring constant $k_2$. The position of the rigid body $x_B$ is described by the position of the first fixed point. Although the system dynamics is characterized by both the kinetic energy and the potential energy, we concentrate on the effects of the parameters appearing in the potential energy in this study. The total potential energy of the FKT model is given by

$$V_{FKT} = \frac{1}{2}\sum_{j=1}^{N} k_2[x_j - \{x_B + (j-1)c_2\}]^2 + \frac{1}{2}\sum_{j=1}^{N} k_1(x_{j\,j+1} - c_2)^2 + \sum_{j=1}^{N} \frac{b}{2}\cos\left(\frac{2\pi x_j}{c_1}\right). \tag{1}$$

where $x_j$ is the position of the $j$th atom and $x_{j\,j+1}$ is the distance of the $j$th atom and the $(j+1)$th atom. The periodic boundary condition is imposed by enforcing the relations

$$x_{N+1} = x_1 + N c_2, \tag{2}$$

and

$$N c_2 = P c_1, \tag{3}$$

where $P$ is an integer. Using $c_1$ as a length scale and $k_1 c_1^2$ as an energy scale, the total potential can be expressed in a dimensionless form given by

$$\tilde{V}_{FKT} = \frac{V_{FKT}}{k_1 c_1^2} = \frac{1}{2}\sum_{j=1}^{N} \tilde{k}_2[\tilde{x}_j - \{\tilde{x}_B + (j-1)\tilde{c}\}]^2 + \frac{1}{2}\sum_{j=1}^{N} (\tilde{x}_{j\,j+1} - \tilde{c})^2 + \sum_{j=1}^{N} \frac{\tilde{b}}{2}\cos(2\pi \tilde{x}_j) \tag{4}$$

where

$$\tilde{x}_j = \frac{x_j}{c_1} \text{ and } \tilde{x}_B = \frac{x_B}{c_1} \tag{5}$$

$$\tilde{k}_2 = \frac{k_2}{k_1}, \tag{6}$$

$$\tilde{c} = \frac{c_2}{c_1} = \frac{P}{N}, \tag{7}$$

and

$$\tilde{b} = \frac{b}{k_1 c_1^2}. \tag{8}$$

Then, the dimensionless friction force $\tilde{F}_R$ is defined as

$$\tilde{F}_R = \frac{F_R}{k_1 c_1}. \tag{9}$$

Thus, the dynamics of the FKT model can be described with three dimensionless parameters $\tilde{k}_2$, $\tilde{c}$ and $\tilde{b}$. When $\tilde{c}$ is not expressed as a ratio of two integers shown in Eq. (7), the model represents an infinite incommensurate system.

One thing to note in the FKT model with respect to frictionless sliding is the existence of the critical surface corrugation $b_c^K$. Below $b_c^K$ there is only one stable state and the kinetic friction becomes zero in the zero-velocity limit, but above $b_c^K$ there are multiple states so that the system exhibits stick-slip motion and transitions between these states result in energy dissipation.

Although the FKT model introduces more degrees of freedom than the Tomlinson model and the FK model, it remains a gross simplification of an AFM experiment. First, in the FKT model the interaction with a surface is described by a fixed one-dimensional periodic potential, but the realistic tip and substrate mate at a more complicated possibly rough two-dimensional interface. The atoms in the substrate are not fixed, but vibrate and change configuration as the result of the interaction with the tip atoms. Secondly, the FKT model has only one layer consisting of a deformable chain of atoms, but a realistic system includes many layers in which atoms are allowed to move. Thirdly, the system in the FKT model is driven by imposing a constant velocity or a constant force to the rigid body, but the real AFM tip is driven through the elastic deformation of the cantilever the other end of which moves at a constant velocity. It is common in simulation to represent these elastic bodies as a spring.

## B. MD simulation model

In this work, we modeled AFM experiments with incommensurate surfaces using a molecular dynamics methodology. Figure 2 (a) illustrates an AFM experiment and Fig. 2 (b) shows the AFM system used for MD simulations. The system models a tip which consists of 263 atoms shown in red and a substrate which consists of 1,800 atoms shown in blue. The substrate is an FCC crystal modeled by the standard Lennard-Jones potential,

$$V(r) = 4\varepsilon_{ab}\left[\left(\frac{\sigma_{ab}}{r}\right)^{12} - \left(\frac{\sigma_{ab}}{r}\right)^{6}\right], \tag{10}$$

where $\varepsilon_{ab}$ is the bond energy between the atom of the type $a$ and the atom of the type $b$, $\sigma_{ab}$ is the characteristic length parameter, and $r$ is the distance between the two atoms. Because the Lennard-Jones potential is short-ranged, we can use a cut-off length to speed computation. The potential is shifted to remove the discontinuity due to the cut-off. We use $\varepsilon$, $\sigma$, and $m$ as the reduced units for energy, length, and mass, respectively. Then, the time unit becomes $\tau = \sigma\sqrt{m/\varepsilon}$. All the quantities are expressed with these units and omitted unless there is ambiguity. The parameters for the substrate are $\sigma_{ss} = 1.0$ and $\varepsilon_{ss} = 1.0$ (the subscript *s* refers to substrate and *t* refers to tip). The cut-off length of between substrate atoms is 2.5 and the cut-off lengths for other pairs are adjusted to yield the identical magnitude of shifted energy. The substrate is subject to periodic boundary conditions in the [110] and [$\bar{1}$10] directions and the size of each periodic cell in these directions is 16.84 × 16.84. The substrate consists of 8 layers in the [001] direction and the atoms on the bottom layer are fixed to prevent a rigid-body translation in the vertical direction.

The tip is created by carving an FCC crystal, which has a larger lattice parameter than the substrate, into a cylindrical shape. The tip is also tapered slightly to the lower side and rotated by 8 degrees about the axis perpendicular to the contact surface. The interactions between the tip atoms are modeled by the harmonic potential, which does not allow any bond breaking, in order to maintain the shape of the tip and prevent wear during sliding. The potential is of the form

$$V(r) = \frac{1}{2} k_{ab} (r - r_{ab})^2 , \qquad (11)$$

where $k_{ab}$ is the stiffness of the atomic bond, and $r_{ab}$ is the equilibrium bond length ($r_{tt} = 1.68$). The tip and the substrate are joined in the [001] direction. The mismatch of the lattice parameters and the rotation prevent the tip atoms from arranging themselves into the lattice of the substrate when the tip atoms are more tightly bound than the interaction between the tip and the substrate. The tip compliances are varied by changing the stiffness $k_{tt}$ of the harmonic potential (hereafter we use $k$ to refer to $k_{tt}$). The interaction between the substrate atoms and the tip atoms are modeled by the Lennard-Jones potential and the parameters are $\sigma_{st} = 0.8$ and $\varepsilon_{st} = 0.5$.

A spring with a stiffness of 10 is linked to the top layer of the tip, which consists of atoms moving together without changing their relative positions, and pulls the tip as the other end of the spring (the slider) moves at a constant speed. The spring stiffness represents the stiffness of the AFM tip and cantilever. A normal force $F_N$ is applied on the top layer of the tip. The sign of the normal force is positive when it is exerted downward. As in Socoliuc's experiments,[6] the magnitudes of the normal force are varied and we observe the dependence of superlubricity on the normal load.

The total potential of this model is given by

$$V_{MD} = \frac{1}{2}k_S(x_S - x_B)^2 + \frac{1}{2}k_S y_B^2 + F_N z_B + \frac{1}{2}\sum_{i>j, i\in t, j\in t} k(r_{ij} - r_{tt})^2$$
$$+ \sum_{i>j, i\in s, j\in s} 4\varepsilon_{ss}\left[\left(\frac{\sigma_{ss}}{r_{ij}}\right)^{12} - \left(\frac{\sigma_{ss}}{r_{ij}}\right)^{6}\right] + \sum_{i\in s, j\in t} 4\varepsilon_{st}\left[\left(\frac{\sigma_{st}}{r_{ij}}\right)^{12} - \left(\frac{\sigma_{st}}{r_{ij}}\right)^{6}\right], \quad (12)$$

where $k_S$ is the stiffness of the spring driving the system, $x_S$ is the slider position, and $(x_B, y_B, z_B)$ is the position vector of the top layer of the tip corresponding to the rigid body in the FKT model. Using $\sigma_{ss}$ as a length scale and $\varepsilon_{ss}$ as an energy scale, the potential can be expressed in the following dimensionless form.

$$\tilde{V}_{MD} = \frac{V_{MD}}{\varepsilon_{ss}} = \frac{1}{2}\tilde{k}_S(\tilde{x}_S - \tilde{x}_B)^2 + \frac{1}{2}\tilde{k}_S \tilde{y}_B^2 + \tilde{F}_N \tilde{z}_B + \frac{1}{2}\sum_{i>j, i\in t, j\in t} \tilde{k}(\tilde{r}_{ij} - \tilde{r}_{tt})^2$$
$$+ \sum_{i>j, i\in s, j\in s} 4\left[\left(\frac{1}{\tilde{r}_{ij}}\right)^{12} - \left(\frac{1}{\tilde{r}_{ij}}\right)^{6}\right] + \sum_{i\in s, j\in t} 4\tilde{\varepsilon}_{st}\left[\left(\frac{\tilde{c}_{st}}{\tilde{r}_{ij}}\right)^{12} - \left(\frac{\tilde{c}_{st}}{\tilde{r}_{ij}}\right)^{6}\right]. \quad (13)$$

The definition of the dimensionless length variables is straightforward and the other dimensionless parameters are defined as follows

$$\tilde{k}_S = \frac{k_S \sigma_{ss}^2}{\varepsilon_{ss}}, \quad (14)$$

$$\tilde{k} = \frac{k \sigma_{ss}^2}{\varepsilon_{ss}}, \quad (15)$$

$$\tilde{\varepsilon}_{st} = \frac{\varepsilon_{st}}{\varepsilon_{ss}}, \quad (16)$$

$$\tilde{c}_{st} = \frac{\sigma_{st}}{\sigma_{ss}}, \quad (17)$$

and

$$\tilde{F}_N = \frac{F_N \sigma_{ss}}{\varepsilon_{ss}}. \quad (18)$$

Then, the system behavior can be described with these dimensionless parameters. In this study, $\tilde{k}$ and $\tilde{F}_N$ have been varied and the other parameters have remained constant ($\tilde{k}_S = 10$, $\tilde{\varepsilon}_{st} = 0.5$, $\tilde{c}_{st} = 0.8$).

All the simulations have been performed at constant temperature ($T = 0.01\,\varepsilon/k_B$) and the temperature is controlled by a variant of the Nose-Hoover thermostat.[17] All the atoms except in the bottom layer of the substrate and the top layer of the tip are subject to the thermostat. With the softest tip we prepared, the velocity dependence was tested at three sliding velocities ($v_S = 10^{-4}$, $10^{-5}$, $10^{-6}$), and all the other tips were simulated at a sliding velocity of $10^{-4}$. The equations of motion are numerically integrated using a modified Velocity-Verlet algorithm.[18]

## III. RESULTS

### A. Friction force and tip compliance

We begin by varying the stiffness values between the tip atoms, $k$, to measure the dependence of the friction force on the tip compliance. This dependence has also been studied in the FKT model where $b_c^K/(k_1 c_1^2)$ is a monotonically increasing function of $k_2/k_1$. When $k_1 = k_2$ with fixed geometry, by dimensional analysis we obtain the relation, $b_c^K \sim k_1 c_1^2$. Therefore, as the interaction between the atoms becomes weaker ($k_1$ decreases), frictionless sliding becomes difficult to obtain because it is possible only in a system with smaller energy corrugation. The simulation study mentioned above [9] also investigated the effects of the stiffness between atoms comprising a surface on the existence of static friction. That study employed flat walls and measured the mean-squared displacement of the top wall, not the friction forces.

Here we measure the friction force directly. Figure 3, which shows only the top layer of the substrate and the lowest layer of the tip, illustrates the initial interface configurations with the stiffest ($k = 100$) and the softest tip ($k = 3.81$). In case of the rigid tip shown in Fig. 3 (a), the tip atoms maintain their original crystalline configuration and the orientation of the tip is also apparent. The atoms of the soft tip conform to the lattice points of the substrate as seen in Fig. 3 (b).

The effect of the arrangement of tip atoms on the lateral force due to the tip compliance is shown in Fig. 4. All the data in this figure were obtained from the simulations performed at a normal force of 5.0 and a sliding velocity of $10^{-4}$. The lateral force $F_R$ is measured by the deformation of the spring as in an AFM experiment. The measured lateral forces are characterized by the continuous, linear increases, followed by relatively quick drops typical of stick-slip motion. We can observe that the softer tips experience a larger lateral force at each transition than the tips with higher stiffnesses. Moreover, while rigid tips show relatively uniform shapes and their lateral forces oscillate around zero, the softer tips display irregular transitions and their lateral forces remain positive after transitions. When $k \leq 10$ as in Fig. 4 (a), we observe many small drops in the lateral force in addition to major decreases. On the contrary, the tips with $k = 15.0$ and $k = 57.2$ shown in Fig. 4 (b) do not show such behavior. These abrupt drops in the lateral force, which also correspond to the abrupt changes in the tip position, imply that transitions here occurred between states defined by local potential energy minima and their neighborhood in configuration space. The existence and stability of multiple meta-stable states causing these small transitions will be discussed later.

Moreover, we simulated ten samples with different initial conditions to check the system-to-system variation. To prepare the samples, we performed simulations with the slider position fixed at zero. During this simulation the positions and velocities of the atoms are periodically saved to generate samples. Therefore, both the initial positions and velocities are different among the samples. With the same stiffness value, all samples showed a similar trend. We averaged the lateral force over the slider position. For a rigid tip ($k = 57.2$), the average friction forces range from 0.08 to 0.1 among the samples while a softer tip ($k = 3.81$) shows a larger variation from 7.34 ~ 8.53.

Figure 5 shows the average lateral force as a function of the stiffness. The error bars show the standard deviation of the samples. Instead of exhibiting monotonically increasing force with decreasing stiffness, we see an abrupt increase around a stiffness of 7. Stiffnesses larger than 7 result in superlubric behavior. The breakdown of superlubricity with softer tips is clearly evident.

### B. Friction force and normal force

The relationship between the friction force and the normal load has been widely investigated in both macroscopic and microscopic friction. As mentioned above, Socoliuc *et. al.*[6] observed that the transition from the vanishing friction to the finite friction due to the stick-slip motion can be controlled by varying the normal force. In this paper we reproduce this dependence of frictionless sliding on the normal force using our simulation model and also present the results at various tip stiffnesses ($k = 10, 15, 25.4, 57.2$). All the simulations with these stiffnesses at a normal force of 5 showed vanishing friction forces in the previous section. It will turn out that the system exhibits different dependence on the normal load at various tip stiffnesses.

First, the adhesion energy and the pull-off force were measured by varying the normal force. The adhesion energy is measured as the total interaction energy between the tip atoms and the substrate atoms, and the pull-off force is determined as the force when the adhesion energy vanishes. As shown in Fig. 6, the adhesion energy increases as the tip becomes rigid and the normal force decreases. However, there is little difference in the pull-off forces at various tip stiffnesses. Therefore, we can conclude that the pull-off force has little dependence on the tip compliance.

Figure 7 shows the lateral force as a function of the slider position at $k = 10$ and at various normal loads. At $F_N = -40$ (the normal force acting upward), there is no sharp drop in the lateral force.

Instead, the lateral force fluctuates around zero continuously with the same periodicity as the lattice parameter of the substrate. This regime corresponds to the condition where only one global minimum exists in the FKT model so that the kinetic friction is negligible. As the normal force increases ($F_N = -20$ and $-10$), sharp drops separated by the substrate lattice parameter are observed and the shape of the curves becomes complex with small bumps. At a normal force acting downward ($F_N = +20$ and $+30$), these small bumps also become sharp drops and the system proceeds through many small and large transitions, which appear as drops in the lateral force curve. When the normal force reaches 50, the lateral force curve does not fluctuate around zero anymore and the superlubric behavior completely breaks down.

The lateral forces at $k = 15$ and $k = 25.4$ are shown in Fig. 8 and Fig. 9 respectively. As at $k = 10$, the lateral forces show continuous sliding without abrupt shifts in the lateral force at lower (upward) normal force. As the normal force rises, multiple states start to emerge and the number of sharp drops increases. Eventually, at the large normal forces of 127 ($k = 15$) and 207 ($k = 25.4$), the lateral forces remain positive at most slider positions. It is also observed that the critical normal forces at which the sharp drops appear and the multiple states emerges are larger at $k = 15$ and $k = 25.4$ than at $k = 10$.

Figure 10 shows the lateral forces at $k = 57.2$. The lateral force exhibits continuous sliding at lower normal load, and as the normal force increases, the stick-slip motion is observed and the multiple states emerge. Unlike the other simulations with smaller tip stiffnesses which simultaneously exhibit the transition from the continuous sliding to stick-slip motion and the emergence of multiple meta-stable states, the simulation data at $k = 57.2$ shows that the sharp drops in the lateral force separated by the substrate lattice parameter (major transitions) develop prior to the emergence of other smaller states and these two regimes are clearly distinguished. Eventually at the normal force of $F_N = 328$ the superlubricity completely breaks down.

Finally, in Fig. 11 we plotted the average lateral force obtained by averaging over the slider position as a function of the normal force. In this graph the average lateral force is rescaled with the tip stiffness and the normal force is shifted by $F_D$ and rescaled with $k$. This rescaling is explained below.

As discussed in Sec. II. B, the system can be characterized by five dimensionless parameters; $\tilde{k}_S$, $\tilde{k}$, $\tilde{\varepsilon}_{st}$, $\tilde{c}_{st}$, and $\tilde{F}_N$. However, since we vary only two parameters $k$ and $F_N$, the system is governed by only these two variables. We can define a dimensionless lateral force $\tilde{F}_R$ as follows

$$\tilde{F}_R \equiv \tilde{k}_S \ (\tilde{x}_S - \tilde{x}_B) = \frac{k_S \sigma_{ss}^2}{\varepsilon_{ss}} (\frac{x_S}{\sigma_{ss}} - \frac{x_B}{\sigma_{ss}}) = \frac{k_S \sigma_{ss}}{\varepsilon_{ss}} (x_S - x_B) = \frac{F_R \sigma_{ss}}{\varepsilon_{ss}} \quad , \tag{19}$$

where $F_R = k_S \ (x_S - x_B)$ is the dimensional definition of the lateral force. As we fixed the geometry, we have the following relation,

$$\tilde{F}_R \sim f_1\left(\tilde{k}_S, \tilde{k}, \tilde{\varepsilon}_{st}, \tilde{F}_N\right). \tag{20}$$

Now assume that the influence of $\tilde{k}_S$ on the system is small and that in analogy to the FKT model we can define a surface energy with a dimensionless corrugation $\tilde{b}$, which is a function of $\tilde{F}_N$, $\tilde{k}$ and $\tilde{\varepsilon}_{st}$. In this way, $\tilde{F}_N$ and $\tilde{\varepsilon}_{st}$ can affect the system only through the energy corrugation $\tilde{b}$. The resulting model correspond to a generalized FKT model and we have the following relation

$$\frac{\tilde{F}_R}{\tilde{k}} \sim f_2\left(\frac{\tilde{b}(\tilde{F}_N, \tilde{k}, \tilde{\varepsilon}_{st})}{\tilde{k}}\right). \tag{21}$$

If we further assume that $\tilde{b} \sim [\tilde{F}_N - \tilde{F}_D(\tilde{k})]$, then the relation Eq. (21) becomes

$$\frac{\tilde{F}_R}{\tilde{k}} \sim f_3\left(\frac{\tilde{F}_N - \tilde{F}_D(\tilde{k})}{\tilde{k}}\right), \tag{22}$$

or

$$\frac{F_R}{k\,\sigma_{ss}} \sim f_3\left(\frac{F_N - F_D(k)}{k\,\sigma_{ss}}\right) \tag{23}$$

where $F_D$ is a critical normal force at which the energy corrugation vanishes and we used it as a fitting parameter at various tip stiffness values. The resulting graph is Fig. 11 and the fitting parameters are given in Table I.

In spite of our rough assumptions, the rescaled lateral force, $F_R/k$, can be well described by the rescaled normal force, $(F_N - F_D)/k$. The deviation from the trend at $k = 10$ and $F_N = 50$ and 60 (two points shown in the right of the graph) represents a breakdown of FKT scaling at low $k$ and high $F_N$. Also, striking is the fact that at low normal force at $k = 57.2$ shown in the inset of Fig. 11, the rescaled lateral force linearly increases, hits a plateau, and then exhibits a drop as the rescaled normal force increases. This dip in the lateral force ranging from 2 to 4 in the rescaled units coincides with the emergence of extra meta-stable states ($F_N = 171$). At this normal force the system makes a transition more easily through two smaller energy barriers rather than one large barrier. However, eventually as the normal force increases, the lateral force starts to increase again more rapidly than before. This behavior is repeated at all stiffnesses except for the smallest stiffness, $k = 10$.

### C. Meta-stable states: Existence and stability

The Tomlinson model can be used to explain zero kinetic friction due to the transition from stick-slip motion to continuous sliding.[6] However, in more complex systems with multiple internal degrees of freedom, the potential energy landscape is more complicated than that in the Tomlinson model. In the FKT model the incommensurate system can have zero static friction even if the surface has some corrugation. The conditions for non-zero static friction, vanishing kinetic friction in the zero-velocity limit, and the existence of the meta-stable states are all coincident in the FKT model.[13] Note that as perfectly incommensurate surfaces can be realized only in infinite systems, we could not observe zero static friction in our simulations with a finite contact area. Instead, we analyzed the relationship between zero kinetic friction and the existence of meta-stable states. In the FKT model,[13] when multiple stable states emerge, kinetic friction cannot vanish because it implies transitions amongst these states result in energy dissipation. However, in our simulations we observed small reconfigurations of the tip atoms that do not necessarily result in macroscale changes in the tip position.

The simulation results presented in the previous sections provide strong evidence of the emergence of meta-stable states before the tip makes a major transition. Non-uniform transitions appear to result from different transition mechanisms associated with various intermediate states. To observe these states explicitly, we performed potential energy minimization using a scheme called FIRE.[19] If two configurations are led to different local minima by the minimization, it can be stated that these configurations belong to different states. This analysis reveals that even in the stick phase where the tip deforms elastically the tip atoms at the interface may continually alter their configurations amongst several metastable states. Because these transitions are not accompanied by noticeable changes in the tip position, there is no significant change in the lateral force. We identified 8 states, and these metastable configurations are shown in Fig. 12. Again, only interface atoms are shown, i.e. the top layer of the substrate and the bottom layer of the tip. Boundary atoms that are less tightly bound are more likely to shift position. The time scale for transitions among these metastable states is much shorter than the time scale related to major transitions. The system continually changes interface configuration amongst the metastable states prior to a major transition.

Figure 13 shows the potential energy differences of each of these states from the value of the #1 state as a function of slider position. Because not all the states are stable at all slider positions, the curves appear only when a stable state was identified. Initially, the #1 state is the most stable state and there is only one metastable state (#2). Starting from at a slider position of 0.33, other metastable states start to appear, and after the slider reaches 0.52, the #2 state become more stable than the #1 state. Eventually, the states branching off from the #2 state (#3, #4, #7) become more stable than the states from the #1 state (#5, #6).

To investigate the frictional behavior of soft tips in more detail in connection to these meta-stable states, we have performed further simulations with the softest tip ($k_{tt}$ = 3.81) by reducing the sliding velocity. Figure 14 shows lateral forces as functions of slider position at three different sliding velocities ($v_S = 10^{-4}, 10^{-5}, 10^{-6}$). While the highest velocity results show sporadic distributions of the transition forces, the lower velocity results exhibit the relatively uniform transitions. Figure 15 shows details of transitions at $v_S = 10^{-4}$ and $v_S = 10^{-5}$. Although the transitions at lower sliding velocity appear uniform, the transition to slip is not a single step transition. The system passes through several intermediate states shown in Fig. 15 (b). These states are similar at each transition. In most cases, two or three tip atoms in the right row are relocated into new sites in the sliding direction, and then the others follow resulting in a major slip. At higher sliding velocity ($v_S = 10^{-4}$) the system undergoes a transition through many complicated mechanisms including local shear inside the tip as shown in Fig. 15 (a), and the mechanisms are different at each transition.

The # 8 configuration in Fig. 12 corresponds to one of the intermediate states shown in Fig. 15 (b). As seen in Fig. 13, after a slider position of 1.44, the #8 state becomes most stable. The probability for major transitions increases as the slider position approaches 1.44 and the energy of the state 8 becomes close that of the state 2. At higher sliding speed, it is probable that the sliding is no longer a quasi-static process and the system does not necessarily pass through the # 8 configuration. The sliding at lower speed becomes quasi-static so that the system follows the change in the slider position adiabatically. Therefore, the system exhibits a rather uniform transition although the mechanism includes numerous transitions.

## IV. CONCLUSION

We have performed molecular dynamics simulations modeling an AFM experiment with incommensurate surfaces to investigate the effects of the tip compliance and the applied normal force on superlubricity. These simulations also provided an opportunity to test the generalizability of the FKT model.

First, in simulations at a constant normal force we observed that while rigid tips exhibit vanishing kinetic friction forces, this superlubric behavior breaks down as the tip stiffness reduces. This breakdown of superlubricity is caused by the rearrangement of the tip atoms into conformity with the substrate lattice, making the interface configuration locally commensurate.

Secondly, we performed simulations by varying normal loads at four different tip stiffness values. In general the tips exhibit vanishing friction at low normal loads, but as the normal force increases superlubric behavior breaks down. This transition from superlubricity to larger friction occurs at different normal loads depending upon the tip stiffnesses. Rigid tips require larger normal loads to suppress the superlubricity. Breakdown of superlubricity is accompanied by the emergence of meta-stable states as evidenced by intermediate drops in the lateral force curves.

Moreover, the validity of the FKT model scaling has been tested by assuming that the interface interaction can be modeled by a periodic potential with an energy corrugation linearly proportional to the applied normal force. All the data from various tip stiffness and normal forces collapse onto a single curve over most of the regime investigated. However, at low stiffness and at high normal load the data deviate from the universal curve representing a breakdown of the FKT rescaling. Furthermore, even when the FKT rescaling applies, non-monotonic behavior that coincides with the emergence of meta-stable states is observed, a feature not present in the simple FKT model.

Finally, with a softer tip, we explicitly identified meta-stable states using a minimization scheme. The stabilities of these states depend on the slider position. The relative stabilities of the meta-stable states result in velocity-dependent transition mechanisms.

## ACKNOWLEDGEMENT

We acknowledge support of the NSF program on Materials and Surface Engineering under Grants CMMI-0510163 and CMMI-0926111 and the use of facilities at the University of Michigan Center for Advanced Computing and the Johns Hopkins University Homewood High Performance Compute Cluster.

[1] D. Dowson, *History of Tribology* (Longman, 1979).

[2] R. W. Carpick and M. Salmeron, Chem. Rev. **97**, 1163 (1997).

[3] G. Binnig, C. F. Quate, and Ch. Gerber, Phys. Rev. Lett. **56**, 930 (1986).

[4] M. Hirano, K. Shinjo, R. Kaneko, and Y. Murata, Phys. Rev. Lett. **67**, 2642 (1991)

[5] M. Dienwiebel, G. S. Verhoeven, N. Pradeep, J. W. M. Frenken, J. A. Heimberg, and H. W. Zandbergen, Phys. Rev. Lett. **92**, 126101 (2004).

[6] A. Socoliuc, R. Bennewitz, E. Gnecco, and E. Meyer, Phys. Rev. Lett. **92**, 134301 (2004).

[7] M. Cieplak, E. D. Smith, and M. O. Robbins, Science **265**, 1209 (1994).

[8] G. He, M. H. Müser, and M. O. Robbins, Science **284**, 1650 (1999).

[9] M. H. Müser and M. O. Robbins, Phys. Rev. B **61**, 2335 (2000).

[10] H. J. Kim, W. K. Kim, M. L. Falk, and D. A. Rigney, Tribol. Lett. **28**, 299 (2007).

[11] M. Chandross, C. D. Lorenz, M. J. Stevens, and G. S. Grest, Langmuir **24**, 1240 (2008).

[12] M. J. Brukman, G. Gao, R. J. Nemanich, and J. A. Harrison, J. Phys. Chem. C **112**, 9358 (2008).

[13] M. Weiss and F.-J. Elmer, Phys. Rev. B **53**, 7539 (1996).

[14] O. M. Braun and Y. S. Kivshar, Phys. Rep. **306**, 1 (1998).

[15] G. A. Tomlinson, Philos. Mag. **7**, 905 (1929).

[16] E. Gnecco, R. Bennewitz, T. Gyalog, Ch. Loppacher, M. Bammerlin, E. Meyer, and H.-J. Güntherodt, Phys. Rev. Lett. **84**, 1172 (2000).

[17] G. J. Martyna, M. L. Klein, and M. Tuckerman, J. Chem. Phys. **97**, 2635 (1992).

[18] G. J. Martyna, M. E. Tuckerman, D. J. Tobias, and M. L. Klein, Mol. Phys. **87**, 1117 (1996).

[19] E. Bitzek, P. Koskinen, F. Gähler, M. Moseler, and P. Gumbsch, Phys. Rev. Lett. **97**, 170201 (2006).

TABLE I. Fitting parameter $F_D$ at various tip stiffnesses $k$

| Tip stiffness, $k$ | Fitting parameter, $F_D$ |
|---|---|
| 10 | -56 |
| 15 | -44 |
| 25.4 | -42 |
| 57.2 | -20 |

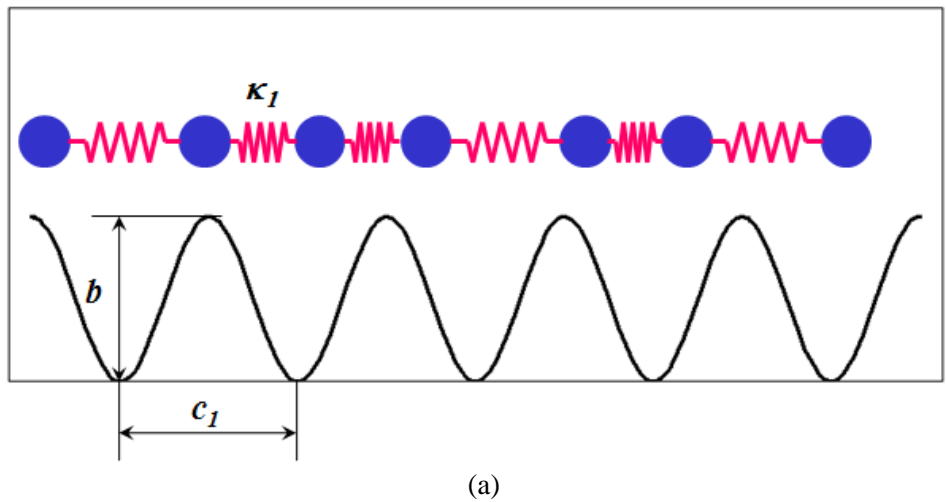

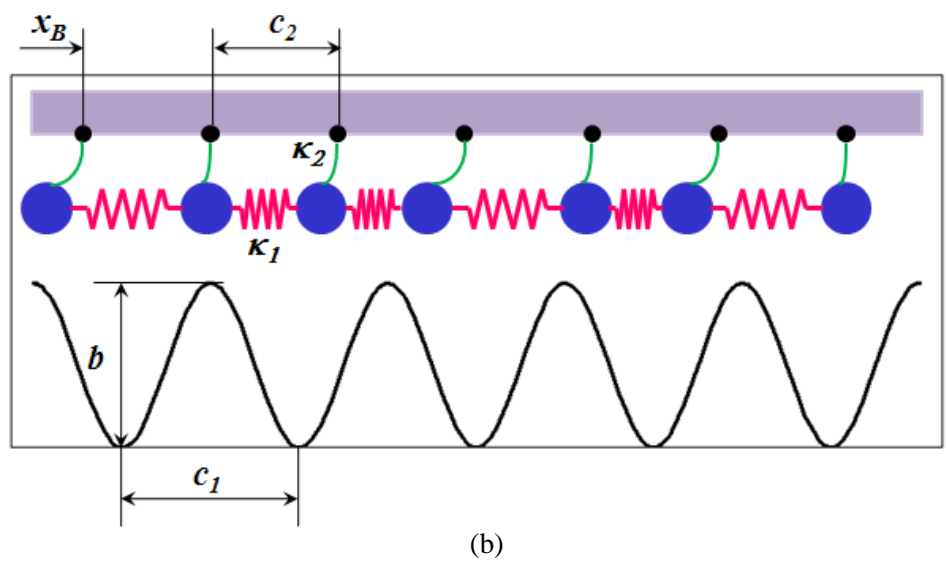

FIG. 1. (a) The Frenkel-Kontorova model. (b) The Frenkel-Kontorova-Tomlinson model.

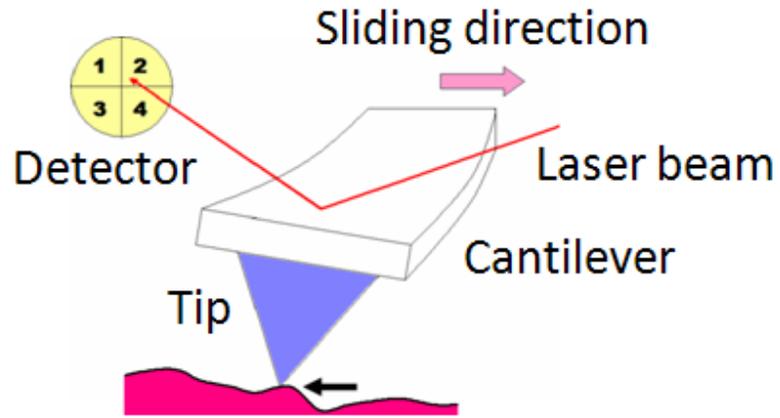

(a)

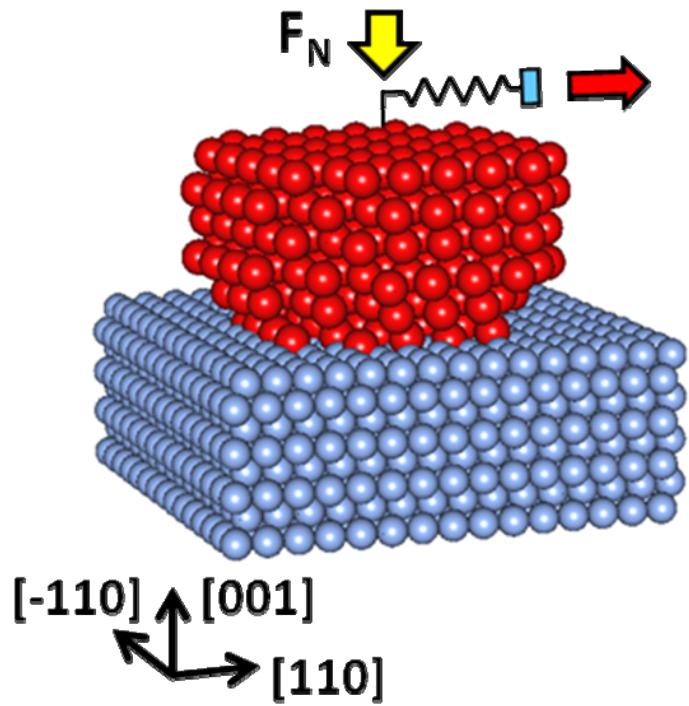

(b)

FIG. 2. (a) An illustration of an AFM experiment. (b) A diagram of 3-dimensional AFM model with a tip and a substrate for MD simulations. The tip atoms are shown in red and the substrate atoms are shown in blue.

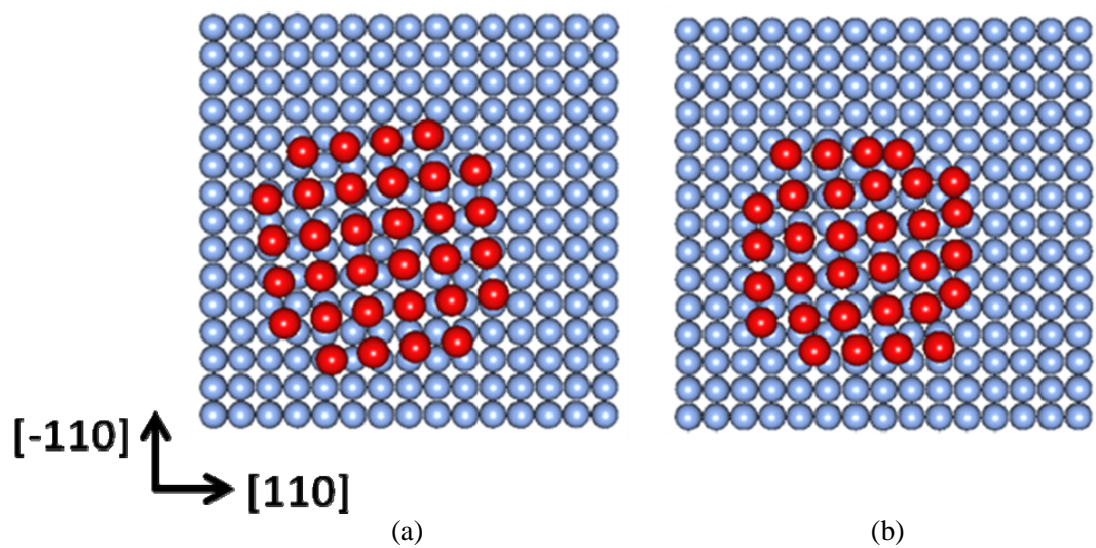

FIG. 3. The configuration of atoms at the interface between the bottom layer of the tip and the top layer of the substrate. (a) A rigid tip; the tip atoms maintain their original crystalline configuration. (b) A softer tip; the tip atoms conform to the lattice of the substrate.

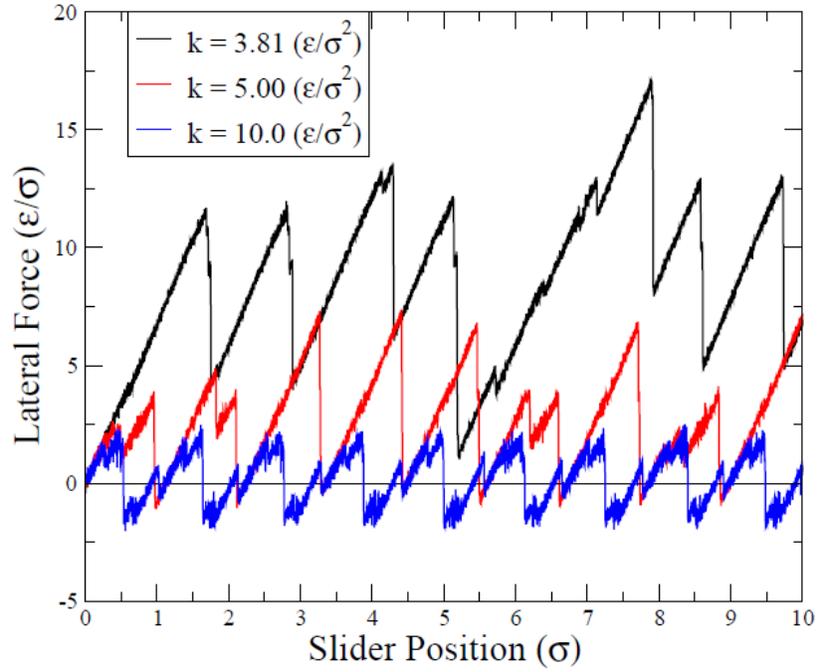

(a)

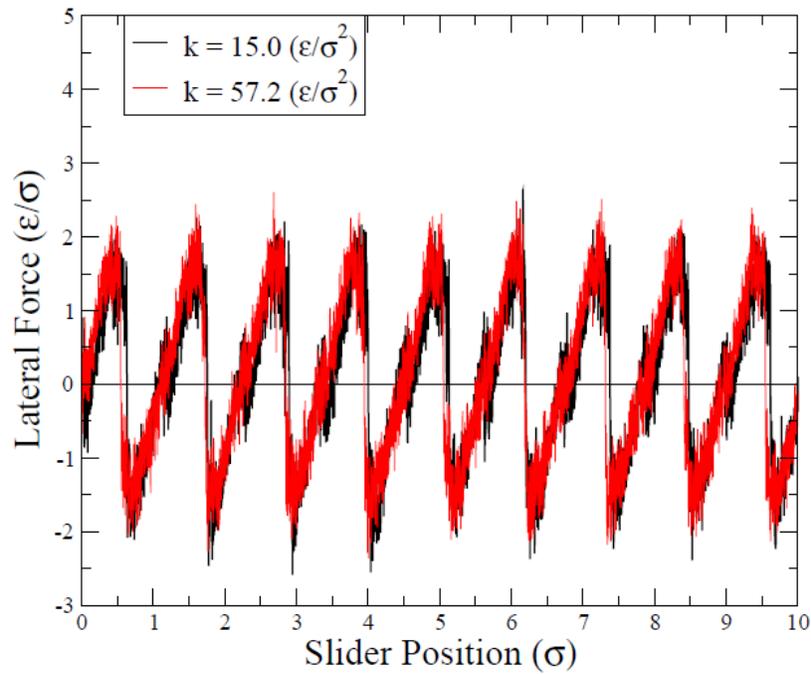

(b)

FIG. 4. Lateral force vs. slider position at different stiffnesses. All the simulations were performed at $F_N=5.0$ and $v_S=10^{-4}$. (a) $k = 3.81, 5.0, 10.0$ ($\varepsilon/\sigma^2$); the softest tip exhibits stick-slip motion and large lateral force and the curves of the other tips have small extra drops. (b) $k = 15.0, 57.2$ ($\varepsilon/\sigma^2$); the lateral forces fluctuate around zero in a uniform way.

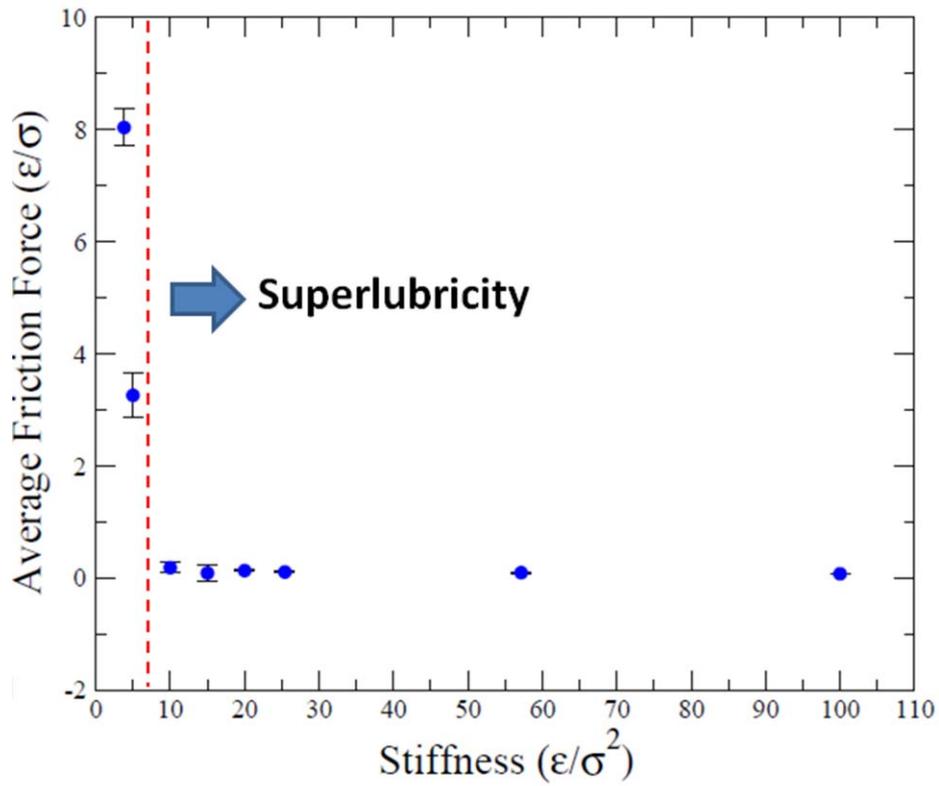

FIG. 5. The relationship between the average lateral force and the tip stiffness. The graph shows two distinct regimes divided by the red dashed line. When the tip stiffness is larger than a critical value (~7) indicated by the red line, superlubricity is observed. When the stiffness is smaller than the critical value, it shows very high force. The error bars show the standard deviation of the samples.

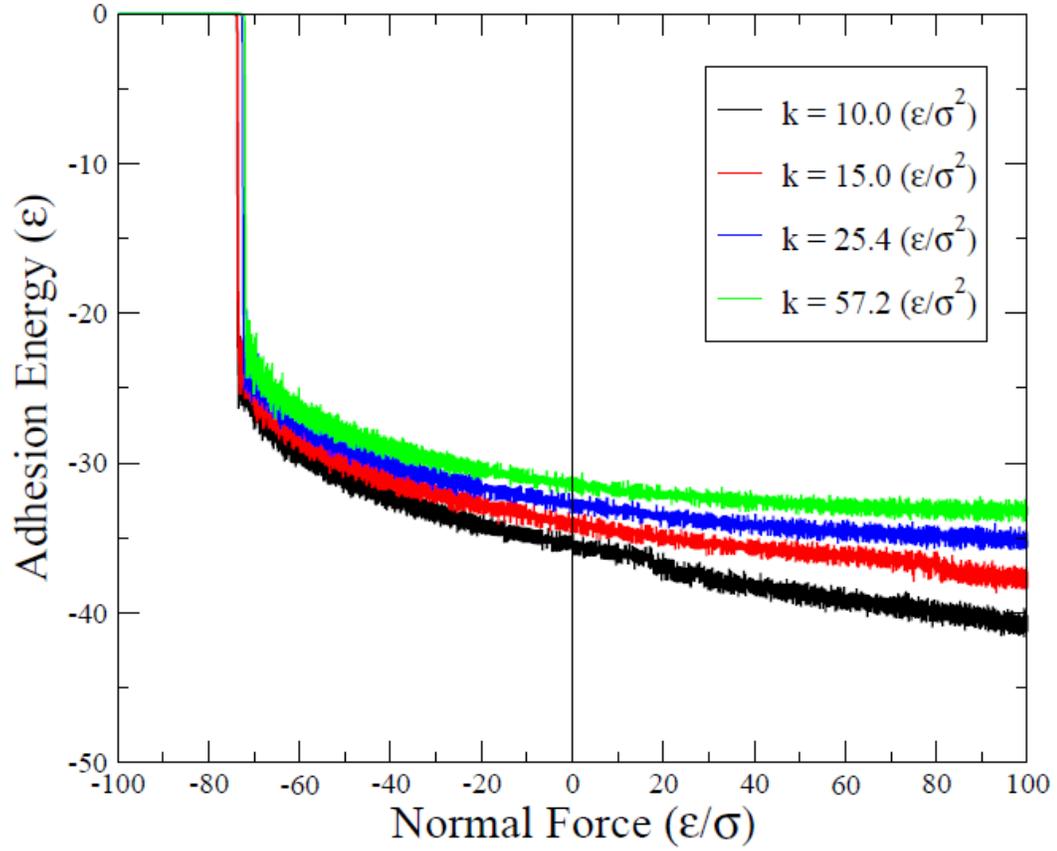

FIG. 6. The adhesion energy between the tip and the substrate as a function of the normal force at various tip stiffnesses ($k$ = 10, 15, 25.4, 57.2). The normal force is positive when it exerts downward. The adhesion energy increases as the tip stiffness rises and the normal force decreases.

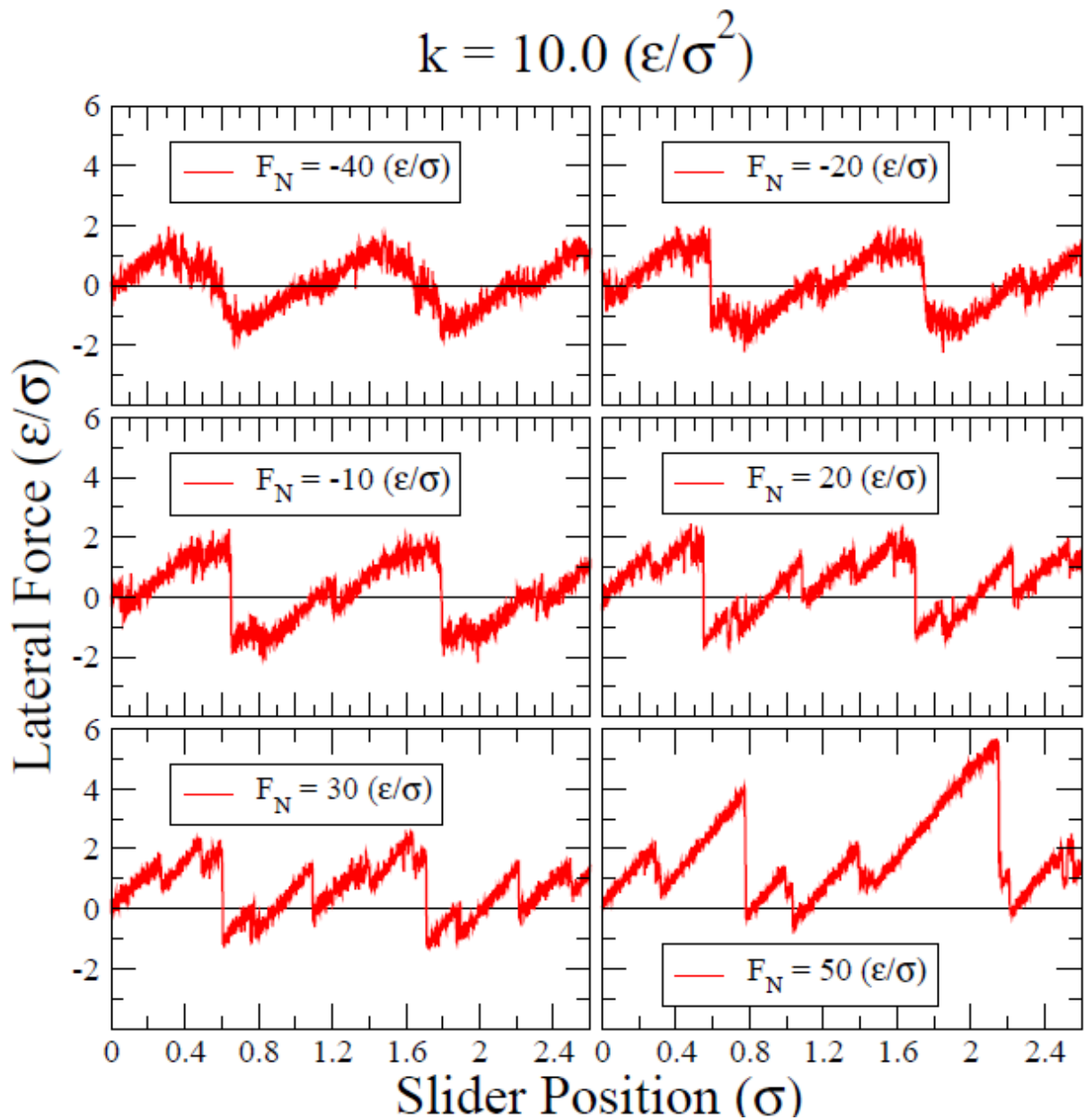

FIG. 7. Lateral force vs. slider position at $k = 10.0$ ($F_N$ = -40, -20, -10, 10, 30, 50). At low normal load the lateral force continuously fluctuate around zero (superlubricity), but as normal force increases several sharp drops appear and eventually friction force becomes large.

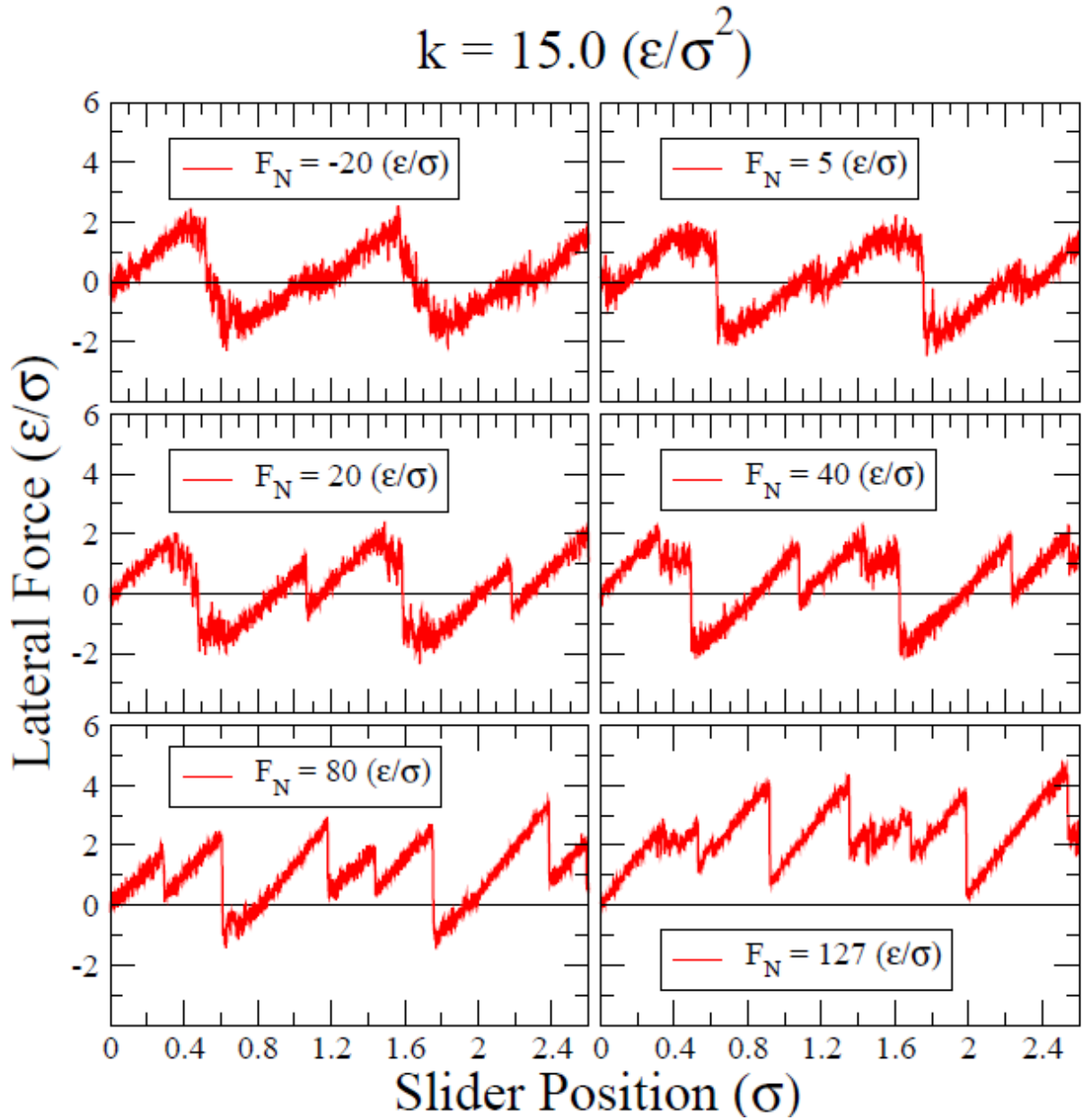

FIG. 8. Lateral force vs. slider position at $k = 15.0$ ($F_N$ = -20, 5, 20, 40, 80, 127). At low normal load the lateral force continuously fluctuate around zero (superlubricity), but as normal force increases several sharp drops appear and eventually friction force becomes large.

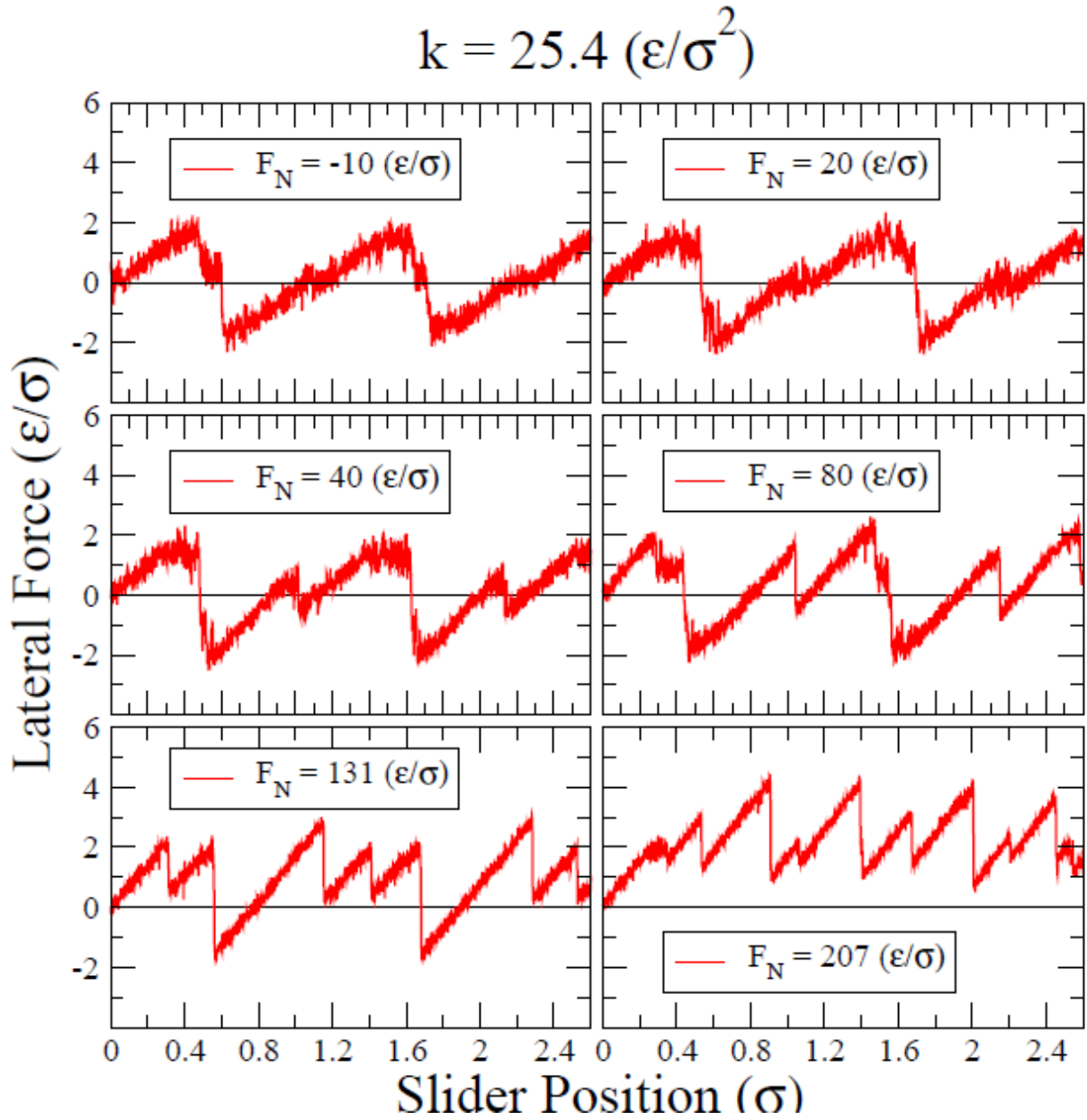

FIG. 9. Lateral force vs. slider position at $k = 25.4$ ($F_N = -10, 20, 40, 80, 131, 207$). At low normal load the lateral force continuously fluctuate around zero (superlubricity), but as normal force increases several sharp drops appear and eventually friction force becomes large.

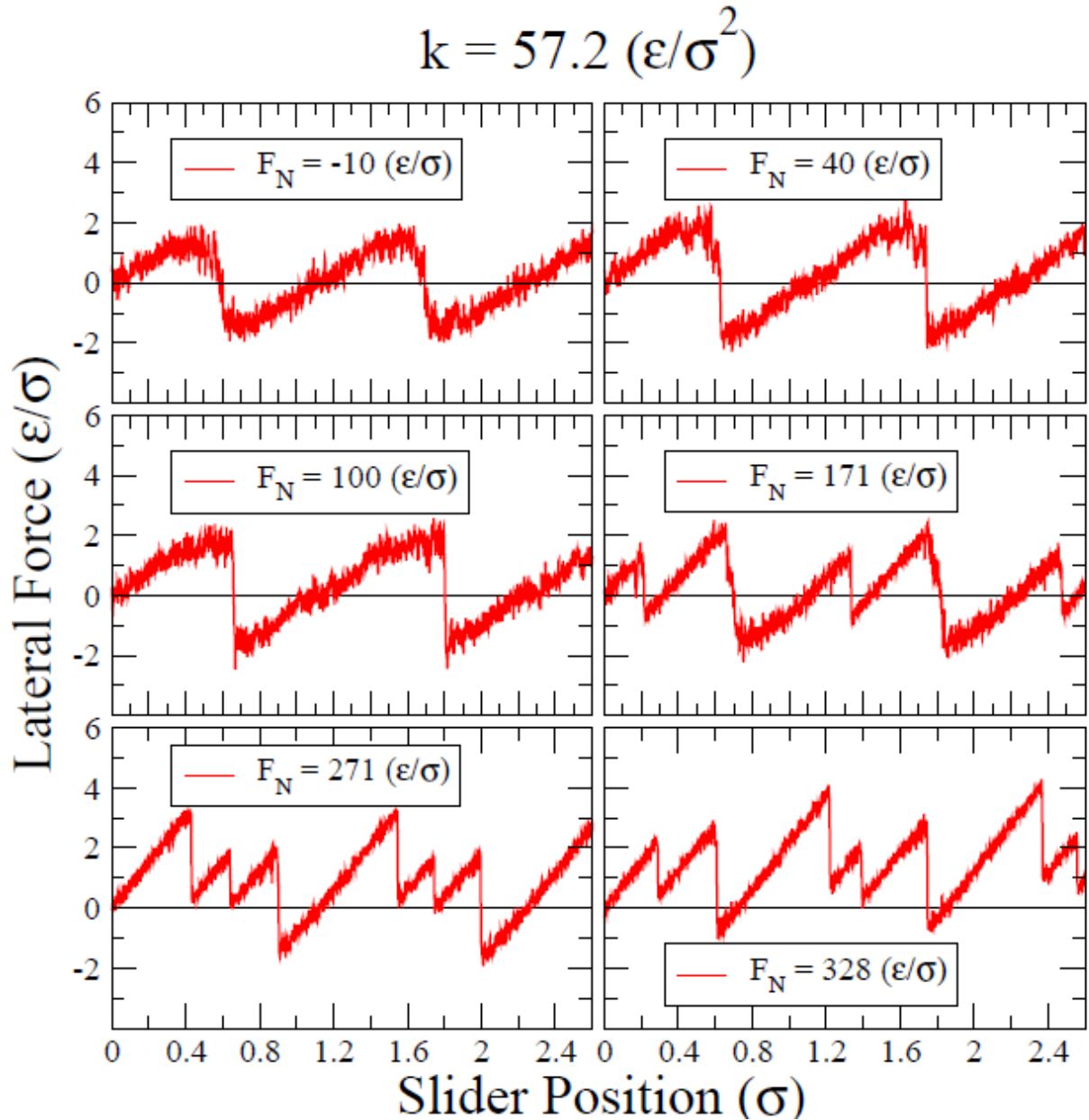

FIG. 10. Lateral force vs. slider position at $k = 57.2$ ($F_N = -10, 40, 100, 171, 271, 328$). At low normal load the lateral force continuously fluctuate around zero (superlubricity), but as normal force increases several sharp drops appear and eventually friction force becomes large. Unlike the other simulations with smaller tip stiffnesses, the sharp drops separated by the substrate lattice parameter develop prior to the emergence of other smaller drops.

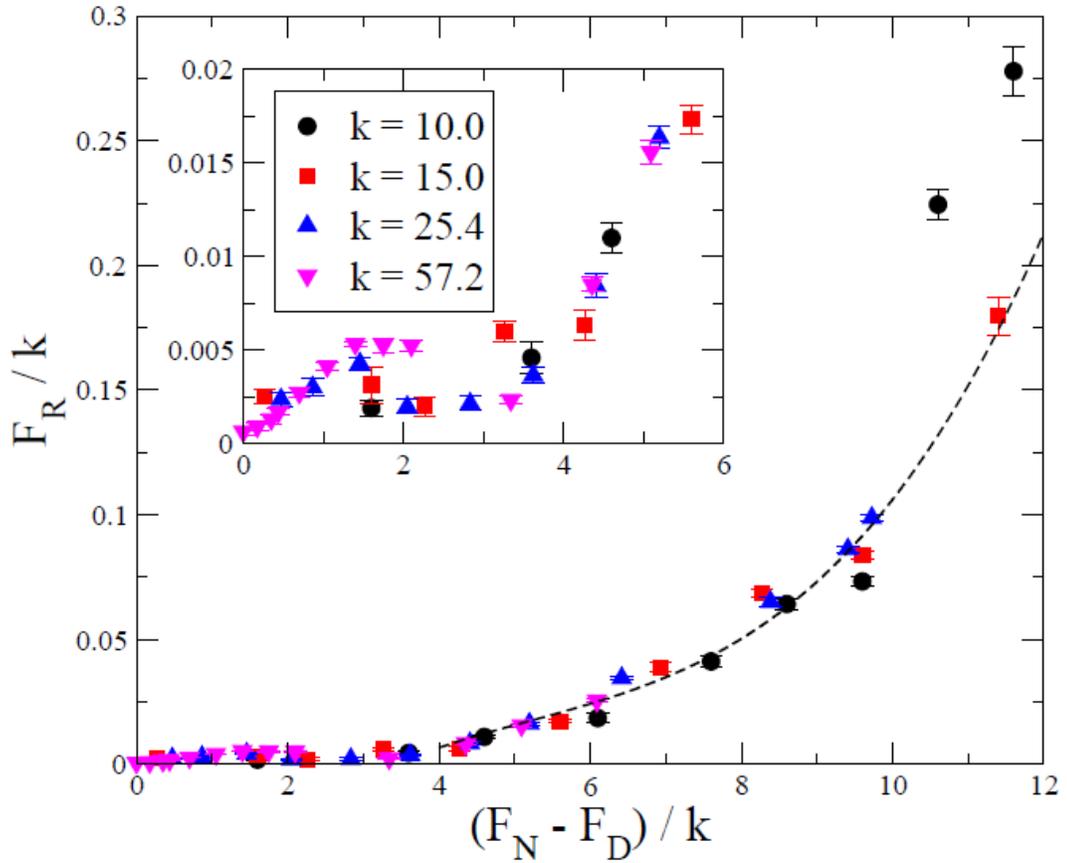

FIG.11. The relationship between the average lateral force and the applied normal load obtained from the simulations with four different tip stiffnesses ($k$ = 10, 15, 25.4, 57.2). Both the lateral force and the normal force are rescaled with the tip stiffness $k$ and the normal force is shifted by $F_D$. The inset shows the details of the data in a smaller range (0 ~ 6) and a dip in the rescaled lateral force is shown. The error bar is the standard deviation and the dashed curve is a trend line of the data points in the intermediate range.

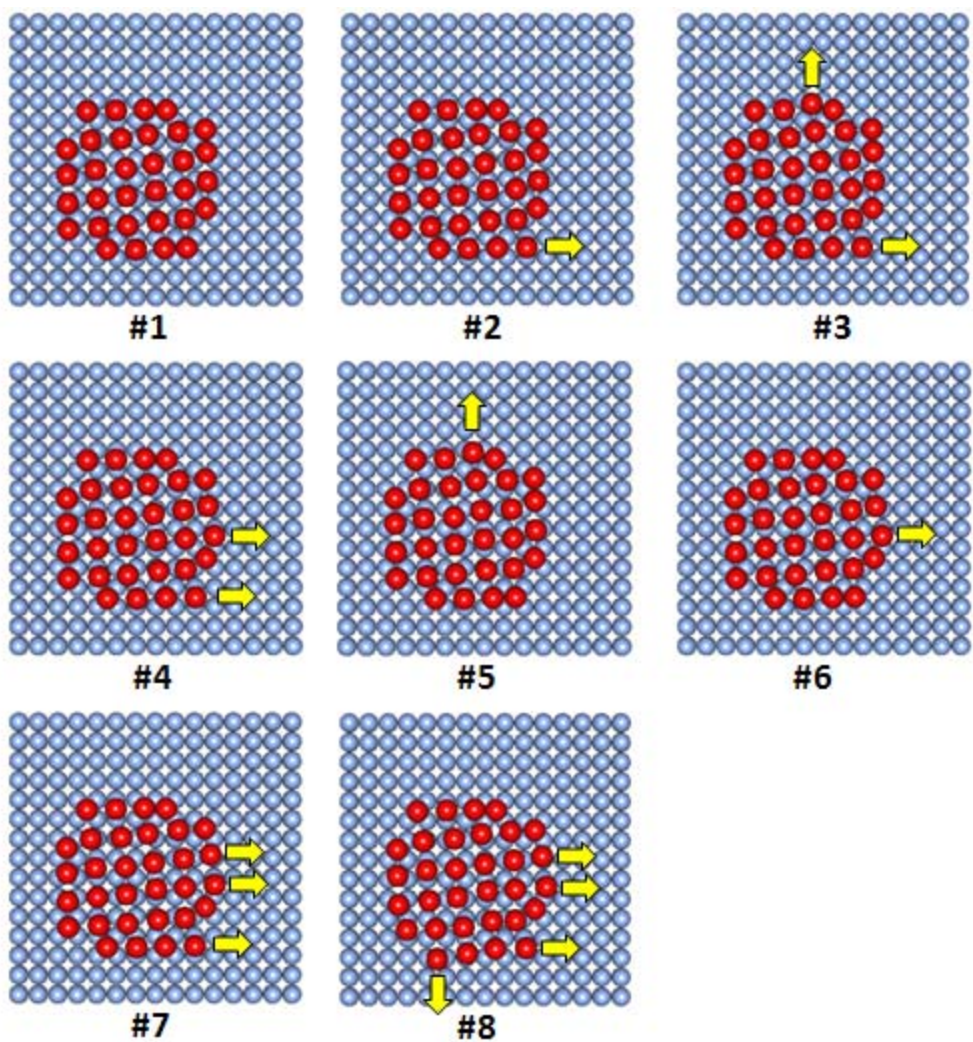

FIG. 12. The metastable states among which the system hops prior to making a transition. Only the two layers at the interface are shown. The yellow arrows indicate the shifts of the atoms relative to the #1 state.

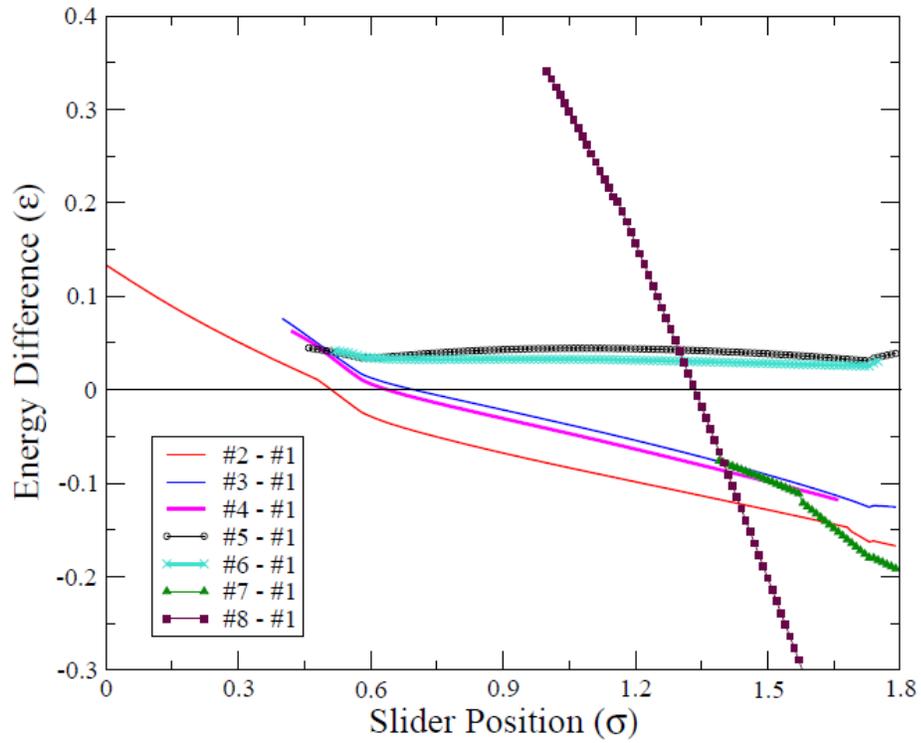

FIG. 13. The energy differences between the metastable states from the #1 state energy as functions of the slider position. Initially the #1 state is the most stable state and there are only 2 stable states (#1 and #2). As the slider advances, other meta-stable states emerge, and after the slider reaches 1.44, the #8 state becomes most stable.

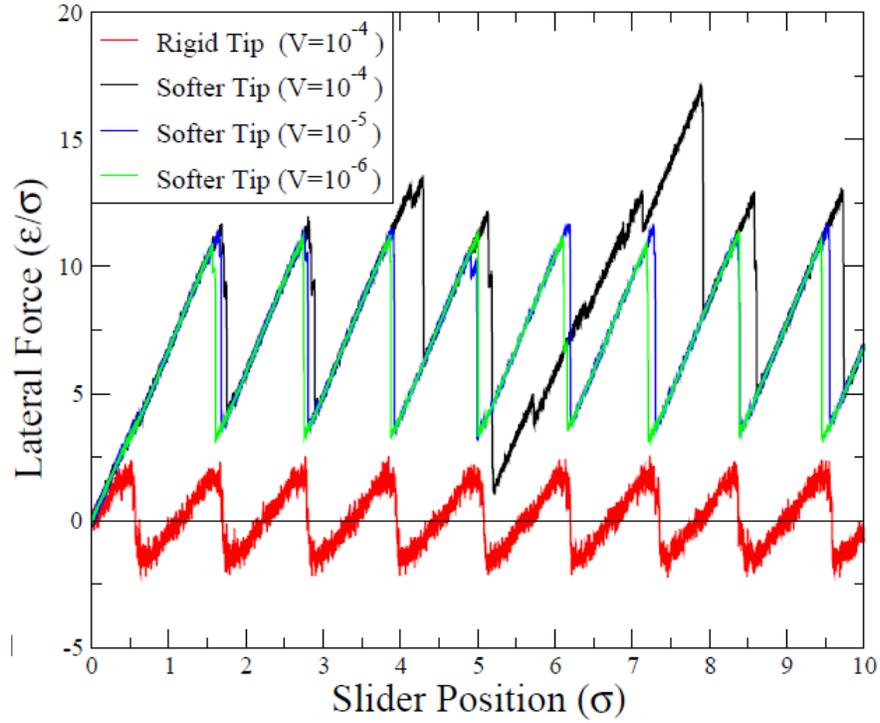

FIG. 14. Lateral Force vs. slider position at different sliding velocities with a rigid tip ($k = 57.2$) and a softer tip ($k = 3.81$). The rigid tip exhibits uniform fluctuation around zero at $v_S = 10^{-4}$, but the soft tip shows sporadic distribution of the forces at higher velocity. At $v_S = 10^{-5}$, the soft tip also shows uniform transitions, but with larger friction forces.

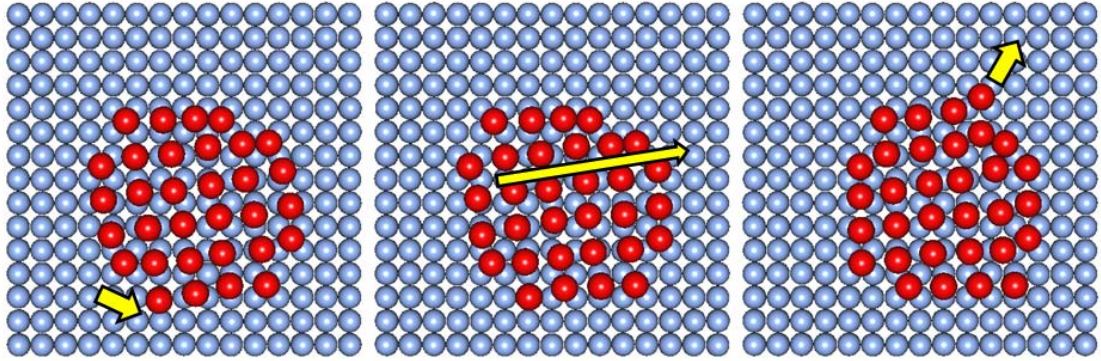

(a)

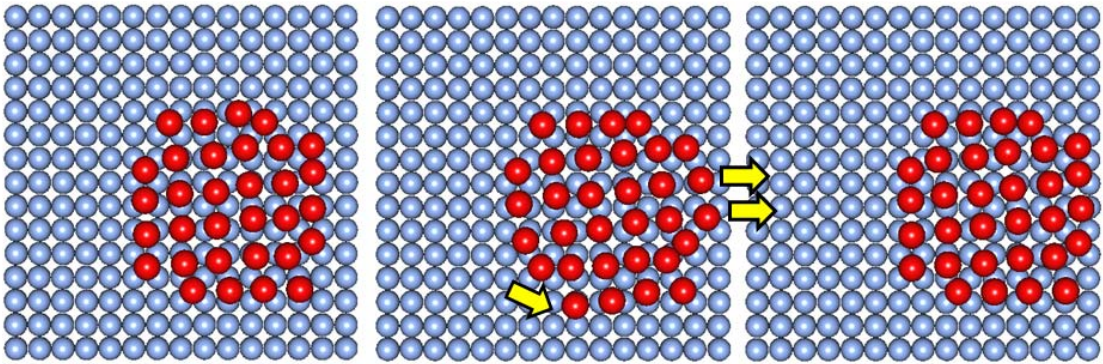

(b)

FIG. 15. Intermediate states the tip passes when making a major transition. (a) $v_S = 10^{-4}$; some atoms move first before the entire tip makes a transition and local shear is also observed. (b) $v_S = 10^{-5}$; two atoms in the right row are relocated into new sites in the sliding direction, and then the other atoms follow resulting in a major slip.